\documentclass[a4paper,12pt]{article}

 \usepackage{amsfonts}
\usepackage{verbatim}

\begin{document}

\begin{flushright} arXiv: 0805.4470 (hep-th)\\ CAS-PHYS-BHU Preprint
\end{flushright} 

\vskip 1cm

\begin{center}

{\Large \bf (Anti-)Dual-BRST Symmetries: Abelian 2-Form Gauge Theory}\\

\vskip 2cm

{\bf R. P. Malik}\\
{\it Centre of Advanced Studies, Physics Department,}\\
{\it Banaras Hindu University, Varanasi-221 005, (U. P.), India}\\
{\small  E-mails: rudra.prakash@hotmail.com ; malik@bhu.ac.in}\\

\end{center}

\vskip 2.5cm

\noindent
{\bf Abstract:}
We derive the absolutely anticommuting (anti-)dual-BRST symmetry transformations
for the appropriate Lagrangian densities of the (3 + 1)-dimensional (4D) free Abelian
2-form gauge theory, under which, the total gauge-fixing term remains invariant. These symmetry
transformations are the analogue of the co-exterior derivative of
differential geometry, in the same sense, as the absolutely anticommuting (anti-)BRST
symmetry transformations are that of the exterior derivative. 
A bosonic symmetry transformation is shown to be the analogue
of the Laplacian operator. The algebraic structures of these symmetry 
transformations are derived and they are demonstrated to be the reminiscent of
the algebra obeyed by the de Rham cohomological operators of differential geometry.\\

              $~~~~~~~~~~~~~~~~~~~~~~~$03.70.+k $~~~$ Theory of quantized fields\\


\newpage

\section{Introduction}

The Becchi-Rouet-Stora-Tyutin (BRST) and anti-BRST symmetry
transformations emerge when the ``classical'' local gauge symmetry transformations
of the local gauge theories are elevated to the 
``quantum'' level. The above (anti-)BRST symmetry transformations are found to be nilpotent
of order two and they anticommute with each-other. These properties are very sacrosanct
and they encode (i) the fermionic nature of these symmetries, and (ii) the linear
independence of these transformations (see, e.g. [1]).

In the realm of the 4D Abelian 2-form [2] gauge theories, the known nilpotent (anti-)BRST
symmetry transformations are found to be anticommuting only up to the U(1)
vector gauge transformations (see, e.g. [3-5]). However, the application of the
superfield approach to BRST formalism, in the context of above 2-form theory, 
requires the above (anti-)BRST symmetry transformations to be
absolutely anticommuting [6]. This key requirement is achieved in
our earlier works [7,8] where a Curci-Ferrari type of restriction is invoked
for the absolute anticommutativity. The above restriction, which happens
to be a key signature of the non-Abelain 1-form gauge theory [9], has been
shown to have a close connection with the concept of gerbes [7].

It is well-known that the (anti-)BRST transformations are the analogue
of the exterior derivative of differential geometry. In our earlier
works [4,5], we have attempted to obtain the symmetry transformations
(for the Abelian 2-form gauge theory) that correspond to the co-exterior
(or dual-exterior) 
derivative and the Laplacian operator of differential geometry. However,
the nilpotent (anti-)dual-BRST symmetry\footnote{These symmetries
would be also called as the (anti-)co-BRST symmetries.} 
transformations (which are the analogue of the
co-exterior derivative) turn out to be anticommuting only up to a U(1) vector
gauge transformation. Thus, the absolute anticommutativity between the 
co-BRST and anti-co-BRST symmetry transformations is absent in the
context of the 4D free Abelain 2-form gauge theory [4,5].
  
In a very recent work [10], we have obtained an appropriate set of
coupled (anti-)BRST invariant Lagrangian densities for the 4D Abelian 2-form
gauge theory. The central purpose of our present investigation is to
generalize the above appropriate Lagrangian densities so as to obtain
the absolutely anticommuting (anti-)BRST as well as the (anti-)co-BRST
symmetry transformations {\it together} for the above Abelian 2-form guage theory. 
Furthermore, we 
obtain a bosonic symmetry transformation (an analogue of the Laplacian operator)
that turns out to be the anticommutator of the (anti-)BRST and (anti-)co-BRST
symmetry transformations. The algebraic structures of the above symmetry transformations
are obtained and they are shown to be the reminiscent of the algebra obeyed by
the de Rham cohomological operators.

The prime factors that have contributed towards our present
investigation are as follows. First and foremost, the known
(anti-)co-BRST symmetry transformations [4,5] are found to be 
non-anticommuting. Thus, it is essential to obtain the correct
anticommuting (anti-)dual-BRST symmetry transformations.
Second, the above symmetry transformations are the essential
ingredients of our theoretical approach to provide a convincing proof
that the 4D 2-form gauge theory is  
a tractable model for the Hodge theory. Our present endeavour 
is a warm-up exercise towards this goal.
Finally, our present understanding
of the 2-form theory would provide useful insights to go
a step further and 
study the higher p-form $(p > 2)$ gauge theories.

The contents of our present paper are organized as follows. In Sec. 2,
we discuss briefly the off-shell nilpotent (anti-)BRST and (anti-)co-BRST
symmetry transformations of our earlier works [4,5] which are found to be
non-anticommuting. Our Sec. 3 is devoted to the derivation of the
absolutely anticommuting (anti-)BRST and (anti-)co-BRST symmetry transformations. In
Sec. 4, we derive a bosonic symmetry transformation.  Our Sec. 5 deals with
the algebraic structures obeyed by the above symmetry transformations and we also
establish their connection with the algebra of the de Rham cohomological operators.
Finally, in Sec. 6, we make some concluding remarks.

\section{Preliminaries: non-anticommuting off-shell nilpotent symmetry transformations}

We begin with the generalized version of the Kalb-Ramond Lagrangian
density (${\cal L}^{(0)} = \frac{1}{12} H^{\mu\nu\kappa} H_{\mu\nu\kappa}$)
for the  4D\footnote{We choose, for the whole body of our present text, 
the 4D flat metric $\eta_{\mu\nu}$ with signature $(+1, -1, -1, -1)$
where the Greek indices $\mu, \nu, \eta...= 0, 1, 2, 3$. The 4D Levi-Civita tensor
$\varepsilon_{\mu\nu\eta\kappa}$ is such that $\varepsilon_{0123} = + 1 =
- \varepsilon^{0123}$ and it
obeys $\varepsilon_{\mu\nu\eta\kappa}
\varepsilon^{\mu\nu\eta\kappa} = - 4!, \varepsilon_{\mu\nu\eta\kappa}
\varepsilon^{\mu\nu\eta\sigma} = - 3! \delta^{\sigma}_{\kappa}$, etc. 
The convention $(\delta B_{\mu\nu}/ \delta B_{\eta\kappa})
= \frac{1}{2!} (\delta_{\mu\eta} \delta_{\nu\kappa} - \delta_{\mu\kappa} \delta_{\nu\eta})$
has been adopted in [4,5].}
free Abelian 2-form gauge theory that respects the off-shell nilpotent
(anti-)BRST and (anti-)dual-BRST symmetry transformations [4,5]. This Lagrangian density is [4,5]
\begin{eqnarray}
&&{\cal L}^{(0)}_B = \frac{1}{2} {\cal B} \cdot {\cal B} - {\cal B}^\mu (\frac{1}{3!} 
\varepsilon_{\mu\nu\eta\kappa} H^{\nu\eta\kappa} - \partial_\mu \phi_2)
 + B^\mu (\partial^\nu B_{\nu\mu} - \partial_\mu \phi_1) - \frac{1}{2} B \cdot B
\nonumber\\ &&- \partial_\mu \bar\beta \partial^\mu \beta
+ (\partial_\mu \bar C_\nu - \partial_\nu \bar C_\mu) (\partial^\mu C^\nu) +
\rho (\partial \cdot C + \lambda) + (\partial \cdot \bar C + \rho) \lambda,
\end{eqnarray}
where $B_\mu = \partial^\nu B_{\nu\mu} - \partial_\mu \phi_1$ and 
${\cal B}_\mu = \frac{1}{3!} \varepsilon_{\mu\nu\eta\kappa} H^{\nu\eta\kappa}
- \partial_\mu \phi_2$ are the  
Lorentz vector auxiliary fields that have been invoked
to linearize the gauge-fixing and kinetic terms, the massless ($\Box \phi_1 =
\Box \phi_2 = 0$) scalar fields $\phi_1$ and $\phi_2$ have been inrtoduced
for the stage-one reducibility in the theory and the totally antisymmetric curvature
tensor $H_{\mu\nu\kappa} = \partial_\mu B_{\nu\kappa} + \partial_\nu B_{\kappa\mu}
+ \partial_\kappa B_{\mu\nu}$ is constructed with the 2-form 
antisymmetric gauge field $B_{\mu\nu}$.

The fermionic Lorentz vector (anti-)ghost fields
$(\bar C_\mu)C_\mu$ (carrying ghost number $(-1)1$) have been introduced 
to compensate for the above gauge-fixing term and they
play important roles in the existence of the 
(anti-)BRST symmetry transformations for the 2-form gauge potential.
The bosonic (anti-)ghost fields
$(\bar\beta)\beta$ (carrying ghost numbers $(-2)2$) are needed for
the requirement of 
ghost-for-ghost in the theory. The auxiliary ghost fields 
$\rho = - \frac{1}{2} (\partial \cdot \bar C)$ and $\lambda = - \frac{1}{2}
(\partial \cdot  C)$ (with ghost numbers (-1)1) are also present in the theory.

The following off-shell nilpotent ($\tilde s_{(a)b}^2 = 0$) (anti-)BRST
symmetry transformations $\tilde s_{(a)b}$\footnote{We adopt
here the standard notations used in our earlier work [7].} for the fields of the Lagrangian
density (1):
\begin{eqnarray}
&& \tilde s_b B_{\mu\nu} = - (\partial_\mu C_\nu - \partial_\nu C_\mu), \quad 
\tilde s_b C_\mu  = - \partial_\mu \beta, \quad \tilde s_b \bar C_\mu = - B_\mu,
\nonumber\\
&& \tilde s_b \phi_1 = \lambda, \quad \tilde s_b \bar \beta = - \rho, \quad
\tilde s_b [\rho, \lambda, \beta, B_\mu, {\cal B}_\mu, \phi_2, H_{\mu\nu\kappa}] = 0, 
\nonumber\\
&& \tilde s_{ab} B_{\mu\nu} = - (\partial_\mu \bar C_\nu - \partial_\nu \bar C_\mu), \quad 
\tilde s_{ab} \bar C_\mu  = + \partial_\mu \bar \beta, \quad \tilde s_{ab}  C_\mu = + B_\mu,
\nonumber\\
&& \tilde s_{ab} \phi_1 = \rho, \quad \tilde s_{ab}  \beta = - \lambda, \quad
\tilde s_{ab} [\rho, \lambda, \bar \beta, B_\mu, {\cal B}_\mu, \phi_2, H_{\mu\nu\kappa}] = 0, 
\end{eqnarray}
leave the Lagrangian density (1) quasi-invariant as it changes to the total spacetime
derivatives (see, e.g. [7,4,5] for details).

The above Lagrangian density (1) also respects the following off-shell nilpotent
($\tilde s_{(a)d}^2 = 0$) (anti-)dual BRST symmetry
transformations $\tilde s_{(a)d}$ [4]
\begin{eqnarray}
&& \tilde s_d B_{\mu\nu} = - \varepsilon_{\mu\nu\eta\kappa} \partial^\eta \bar C^\kappa, \quad 
\tilde s_d \bar C_\mu  = + \partial_\mu \bar \beta, \quad \tilde s_d  C_\mu = - {\cal B}_\mu,
\nonumber\\
&& \tilde s_d \phi_2 = \rho, \quad \tilde s_d  \beta = - \lambda, \quad
\tilde s_d [\rho, \lambda, \bar \beta, B_\mu, {\cal B}_\mu, \phi_1, (\partial^\nu B_{\nu\mu})] = 0, 
\nonumber\\
&& \tilde s_{ad} B_{\mu\nu} = - \varepsilon_{\mu\nu\eta\kappa} \partial^\eta  C^\kappa, \quad 
\tilde s_{ad}  C_\mu  = - \partial_\mu  \beta, \quad \tilde s_{ad} \bar C_\mu = + {\cal B}_\mu,
\nonumber\\
&& \tilde s_{ad} \phi_2 = \lambda, \quad \tilde s_{ad} \bar \beta = - \rho, \quad
\tilde s_{ad} [\rho, \lambda,  \beta, B_\mu, {\cal B}_\mu, \phi_1, (\partial^\nu B_{\nu\mu})] = 0. 
\end{eqnarray}
It can be readily checked that the Lagrangian density (1) transforms to a total
spacetime derivative under the above off-shell nilpotent transformations $\tilde s_{(a)d}$.

The noteworthy points at this stage are (i) under the (anti-)BRST and 
(anti-)co-BRST transformations, the
curvature tensor $H_{\mu\nu\kappa}$ and the gauge-fixing term $(\partial^\nu B_{\nu\mu})$
remain invariant, respectively, and (ii) the anticommutators of the (anti-)BRST and (anti-)co-BRST transformations produce non-zero results when they act on the fermionic (anti-)ghost fields $(\bar C_\mu)C_\mu$. Both these observations are very important for
our present discussions.

In the language of the cohomological operators, the curvature tensor $H_{\mu\nu\kappa}$ owes
its origin to the exterior derivative $d = dx^\mu \partial_\mu$ (with $d^2 = 0$) because the
3-form $H^{(3)} = \frac{1}{3!} (dx^\mu \wedge dx^\nu \wedge dx^\kappa) H_{\mu\nu\kappa}$ 
defines it through $H^{(3)} = d B^{(2)}$ where $B^{(2)} = \frac{1}{2!} (dx^\mu \wedge dx^\nu) B_{\mu\nu}$ introduces the gauge potential $B_{\mu\nu}$. The operation of the co-exterior derivative
$\delta = - * d *$ (with $\delta^2 = 0$) on the 2-form  produces the gauge-fixing term (i.e. $\delta B^{(2)} = dx^\mu (\partial^\nu B_{\nu\mu})$). Here $*$ is the Hodge duality operation
on the 4D spacetime manifold. Thus, the nilpotent 
(anti-)BRST and (anti-)co-BRST symmetry transformations
owe their origin to $d$ and $\delta$, respectively (because $H_{\mu\nu\kappa}$ and
$(\partial^\nu B_{\nu\mu})$ remain invariant under them).

It can be checked that $(\tilde s_b \tilde s_{ab} + \tilde s_{ab} \tilde s_b) C_\mu 
= \partial_\mu \lambda$ and $(\tilde s_b \tilde s_{ab} + \tilde s_{ab} \tilde s_b) \bar C_\mu 
= - \partial_\mu \rho$. Similarly, the anticommutators 
$\{ \tilde s_d, \tilde s_{ad} \} C_\mu \equiv  (\tilde s_d \tilde s_{ad} 
+ \tilde s_{ad} \tilde s_d) C_\mu 
= \partial_\mu \lambda$ and $\{ \tilde s_d, \tilde s_{ad} \} \bar C_\mu \equiv
(\tilde s_d \tilde s_{ad} + \tilde s_{ad} \tilde s_d) \bar C_\mu 
= - \partial_\mu \rho$ are not equal to zero. The above anticommutators
for the rest of the fields, however, turn out to be absolutely anticommuting.
Thus, we note that the (anti-)BRST and
(anti-)co-BRST symmetry transformations are anticommuting only up to the U(1) vector
gauge transformations. They are {\it not} absolutely anticommuting in nature.

\section{Anticommuting off-shell nilpotent symmetry transformations}

We begin with the appropriate BRST  (${\cal L}_B$) and anti-BRST 
(${\cal L}_{\bar B}$) invariant Lagrangian 
densities that have been proposed in our very recent work [10] connected with the 
4D free Abelian 2-form gauge theory.
These are\footnote{We follow here the convention 
$(\delta B_{\mu\nu}/ \delta B_{\eta\kappa}) = \frac{1}{2!} (\delta_{\mu\eta} \delta_{\nu\kappa}
- \delta_{\mu\kappa} \delta_{\nu\eta})$ which is exactly the same as the one
adopted in [4,5]. It is also clear that $\varepsilon_{\mu\nu\eta\kappa} \partial^\nu B^{\eta\kappa}
= \frac{1}{3} \varepsilon_{\mu\nu\eta\kappa} H^{\nu\eta\kappa}$.} [10]
\begin{eqnarray}
&&{\cal L}_B = \frac{1}{6} H^{\mu\nu\kappa} H_{\mu\nu\kappa} + B^\mu (\partial^\nu B_{\nu\mu}
- \partial_\mu  \phi_1) + B \cdot B + \partial_\mu \bar \beta \partial^\mu \beta \nonumber\\ 
&& + (\partial_\mu \bar C_\nu - \partial_\nu \bar C_\mu) (\partial^\mu C^\nu)
+ (\partial \cdot C - \lambda) \rho + (\partial \cdot \bar C + \rho)\; \lambda, \nonumber\\
&&{\cal L}_{\bar B} = \frac{1}{6} H^{\mu\nu\kappa} H_{\mu\nu\kappa} 
+ \bar B^\mu (\partial^\nu B_{\nu\mu}
+ \partial_\mu \phi_1) + \bar B \cdot \bar B + \partial_\mu \bar \beta \partial^\mu \beta \nonumber\\ 
&& + (\partial_\mu \bar C_\nu - \partial_\nu \bar C_\mu) (\partial^\mu C^\nu)
+ (\partial \cdot C - \lambda) \rho + (\partial \cdot \bar C + \rho)\; \lambda.
\end{eqnarray}
We focus on these Lagrangian densities for the discussion of all the underlying
symmetries of the 4D free Abelian 2-form gauge theory.
The kinetic term $\frac{1}{6} (H^{\mu\nu\kappa} H_{\mu\nu\kappa})$ 
can be linearlized by introducing
the auxiliary Lorentz vector fields ${\cal B}_\mu, \bar {\cal B}_\mu$  and massless 
($ \Box \phi_2 = 0$) scalar field $\phi_2$ as
\begin{eqnarray}
&& {\cal L}_{(B, {\cal B})} = {\cal B} \cdot {\cal B} 
- {\cal B}^\mu (\varepsilon_{\mu\nu\eta\kappa} \partial^\nu B^{\eta\kappa} - \partial_\mu \phi_2) 
+ B^\mu (\partial^\nu B_{\nu\mu} - \partial_\mu \phi_1) + B \cdot B \nonumber\\
&&+ \partial_\mu \bar\beta \partial^\mu \beta  
+ (\partial_\mu \bar C_\nu - \partial_\nu \bar C_\mu) (\partial^\mu C^\nu) 
+ (\partial \cdot C - \lambda) \rho + (\partial \cdot \bar C + \rho)\; \lambda,
\end{eqnarray}
\begin{eqnarray}
&& {\cal L}_{(\bar B, \bar {\cal B})} = \bar {\cal B} \cdot \bar {\cal B} 
- \bar {\cal B}^\mu (\varepsilon_{\mu\nu\eta\kappa} \partial^\nu B^{\eta\kappa} + \partial_\mu \phi_2) 
+ \bar B^\mu (\partial^\nu B_{\nu\mu} + \partial_\mu \phi_1) + \bar B \cdot \bar B \nonumber\\
&&+ \partial_\mu \bar\beta \partial^\mu \beta  
+ (\partial_\mu \bar C_\nu - \partial_\nu \bar C_\mu) (\partial^\mu C^\nu) 
+ (\partial \cdot C - \lambda) \rho + (\partial \cdot \bar C + \rho)\; \lambda.
\end{eqnarray}
Here all the mathematical symbols denote their usual meanings (cf. Sec. 2).

The following off-shell nilpotent ($s_{(a)b}^2 = 0$) and anticommuting
($s_b s_{ab} + s_{ab} s_b = 0$) (anti-)BRST transformations on the fields of (5) and (6), namely;
\begin{eqnarray}
&&s_b B_{\mu\nu} = - (\partial_\mu C_\nu - \partial_\nu C_\mu), \quad s_b C_\mu = - \partial_\mu
\beta, \quad s_b \bar C_\mu = - B_\mu, \nonumber\\
&& s_b \phi_1 = \lambda, \qquad s_b \bar \beta = - \rho, \qquad
s_b [\lambda, \rho, \beta, \phi_2, {\cal B}_\mu, B_\mu, H_{\mu\nu\kappa}] = 0, \nonumber\\
&&s_{ab} B_{\mu\nu} = - (\partial_\mu \bar C_\nu - \partial_\nu \bar C_\mu), \quad 
s_{ab} \bar C_\mu = - \partial_\mu
\bar \beta, \quad s_{ab}  C_\mu = + \bar  B_\mu, \nonumber\\
&& s_{ab} \phi_1 = \rho, \quad s_{ab}  \beta = - \lambda, \quad
s_{ab} [\lambda, \rho, \bar \beta, \phi_2, \bar {\cal B}_\mu, \bar B_\mu, H_{\mu\nu\kappa}] = 0,  
\end{eqnarray}
are the symmetry transformations for (5) and (6) because
\begin{eqnarray}
s_b {\cal L}_{(B, {\cal B})} &=& - \;\partial_\mu \;\bigl [(\partial^\mu C^\nu - \partial^\nu C^\mu) B_\nu 
+ \lambda B^\mu  + \rho \partial^\mu \beta \bigr ], \nonumber\\
s_{ab} {\cal L}_{(\bar B, \bar {\cal B})} &=& -\; \partial_\mu \;\bigl [(\partial^\mu \bar C^\nu 
- \partial^\nu \bar C^\mu) \bar B_\nu 
- \rho \bar B^\mu  + \lambda \partial^\mu \bar \beta \bigr ].
\end{eqnarray}
Thus, the Lagrangian densities change to the spacetime total derivatives.

Similarly, the following off-shell nilpotent ($s_{(a)d}^2 = 0$) and absolutely anticommuting
($s_d s_{ad} + s_{ad} s_d = 0$) (anti-)co-BRST transformations
\begin{eqnarray}
&& s_d B_{\mu\nu} = - \frac{1}{2} \varepsilon_{\mu\nu\eta\kappa} \partial^\eta \bar C^\kappa, \quad 
s_d \bar C_\mu  = + \partial_\mu \bar \beta, \quad s_d  C_\mu = - {\cal B}_\mu,
\nonumber\\
&& s_d \phi_2 = - \rho, \quad s_d  \beta =  \lambda, \quad
s_d [\rho, \lambda, \bar \beta, B_\mu, {\cal B}_\mu, \phi_1, (\partial^\nu B_{\nu\mu})] = 0, 
\nonumber\\
&& s_{ad} B_{\mu\nu} = - \frac{1}{2} \varepsilon_{\mu\nu\eta\kappa} \partial^\eta  C^\kappa, \quad 
s_{ad}  C_\mu  = + \partial_\mu  \beta, \quad  s_{ad} \bar C_\mu = + \bar {\cal B}_\mu,
\nonumber\\
&& s_{ad} \phi_2 = - \lambda, \quad  s_{ad} \bar \beta =  \rho, \quad
s_{ad} [\rho, \lambda,  \beta, \bar B_\mu, 
\bar {\cal B}_\mu, \phi_1, (\partial^\nu B_{\nu\mu})] = 0, 
\end{eqnarray}
leave the Lagrangian densities (5) and (6) quasi-invariant because
\begin{eqnarray}
s_{d} {\cal L}_{(B,  {\cal B})} &=& +\; \partial_\mu \;\bigl [(\partial^\mu \bar C^\nu 
- \partial^\nu \bar C^\mu) {\cal B}_\nu 
- \rho {\cal B}^\mu  + \lambda \partial^\mu \bar \beta \bigr ], \nonumber\\
s_{ad} {\cal L}_{(\bar B,  \bar {\cal B})} &=& + \;\partial_\mu\; \bigl [(\partial^\mu  C^\nu 
- \partial^\nu  C^\mu) \bar {\cal B}_\nu 
+ \lambda \bar {\cal B}^\mu  + \rho \partial^\mu \beta \bigr ].
\end{eqnarray}
The absolutely anticommuting (cf. Sec. 5 below)
and nilpotent (anti-)BRST as well as (anti-)co-BRST transformations
are the {\it symmetry} transformations of the equivalent and coupled 
Lagrangian densities (5) and (6).

Now, the stage is set to comment on the anticommutativity properties of the 
off-shell nilpotent (anti-)BRST
and (anti-)co-BRST symmetry transformations. The equations of motion, that emerge from
(5) and (6), are
\begin{eqnarray}
&& B_\mu = - \frac{1}{2} \bigl (\partial^\nu B_{\nu\mu} - \partial_\mu \phi_1), \quad
\bar B_\mu = - \frac{1}{2} \bigl (\partial^\nu B_{\nu\mu} + \partial_\mu \phi_1), \nonumber\\
&& {\cal B}_\mu =  \frac{1}{2} \bigl (
\varepsilon_{\mu\nu\eta\kappa} \partial^\nu B^{\eta\kappa} - \partial_\mu \phi_2), \quad
\bar {\cal B}_\mu =  \frac{1}{2} \bigl (
\varepsilon_{\mu\nu\eta\kappa} \partial^\nu B^{\eta\kappa} + \partial_\mu \phi_2).
\end{eqnarray}
The above relations imply: $\Box \phi_1 = \Box \phi_2 = 0$ and $\partial \cdot B
= \partial \cdot \bar B = \partial \cdot {\cal B} = \partial \cdot \bar {\cal B} = 0$.
Furthermore, we obtain the Curci-Ferrari type of restrictions
\begin{eqnarray}
B_\mu - \bar B_\mu - \partial_\mu \phi_1 = 0, \; \qquad \;
{\cal B}_\mu - \bar {\cal B}_\mu + \partial_\mu \phi_2 = 0,
\end{eqnarray}
which enable us to prove that $(s_b s_{ab} + s_{ab} s_b) = 0$ and 
$(s_d s_{ad} + s_{ad} s_d) = 0$. In particular, it can be checked that
$\{s_b, s_{ab} \} B_{\mu\nu} = 0$ and $\{s_d, s_{ad} \} B_{\mu\nu} = 0$ only
if we exploit the Curci-Ferrari type restrictions of (12) (cf. Sec. 5 also).

Using nilpotent transformations (7) and (9), it can be explicitly checked
that the Curci-Ferrari type restrictions (12) are the (anti-)BRST as well
as (anti-)co-BRST invariant quantities (see, Sec. 5 below, for more details). 
As a consequence, in some sense, they are very much
``physical'' in nature. Thus, the imposition of these restrictions, in
the proof of anticommutativity of the (anti-)BRST and (anti-)co-BRST
symmetry transformations, is physically {\it not} unsolicited. Furthermore,
it has been shown in our earlier work [7] that one of the above restrictions
(i.e. $B_\mu - \bar B_\mu - \partial_\mu \phi_1 = 0$) is connected with
the geometrical objects called gerbes. Thus, relations (12) are interesting.

\section{Bosonic symmetry transformations}

It is crystal clear that the coupled and equivalent Lagrangian densities (5)
and (6) respect {\it four} nilpotent symmetry transformations. In particular,
the Lagrangian density (5) is endowed with the BRST and dual-BRST symmetry transformations
and the symmetry transformations for the Lagrangian density (6) are the anti-BRST
and anti-dual-BRST transformations.  It is very natural to expect the existence 
of a bosonic
symmetry transformation $s_w = \{ s_b, s_d \}$  (with $s_w^2 \neq 0$) for the
Lagrangian density (5) and $s_{\bar w} = \{ s_{ab}, s_{ad} \}$  (with $s_{\bar w}^2 \neq 0$)
for that of the Lagrangian density (6).

The following local and infinitesimal bosonic transformations
\begin{eqnarray}
&& s_w B_{\mu\nu} = \partial_\mu {\cal B}_\nu - \partial_\nu {\cal B}_\mu + \frac{1}{2}
\varepsilon_{\mu\nu\eta\kappa} \partial^\eta B^\kappa, \quad s_w C_\mu = - \partial_\mu \lambda,
\nonumber\\
&& s_w \bar C_\mu = - \partial_\mu \rho, \quad s_w [\rho, \lambda, \phi_1, \phi_2, \beta, \bar\beta,
B_\mu, {\cal B}_\mu ] = 0, 
\end{eqnarray}
are the {\it symmetry} transformations for the Lagrangian density (5) because
\begin{eqnarray}
s_w {\cal L}_{(B, {\cal B})} &=& \partial_\mu \Bigl [\lambda (\partial^\mu \rho) - 
(\partial^\mu \lambda) \rho + B^\kappa \partial^\mu {\cal B}_\kappa \nonumber\\
 &-& {\cal B}^\kappa \partial^\mu B_\kappa + {\cal B}^\mu (\partial \cdot B) 
- B^\mu (\partial \cdot {\cal B}) \Bigr ]. 
\end{eqnarray}
Thus, we note that the anticommutator $\{s_b, s_d \}$ does generate a bosonic symmetry
transformation for the Lagrangian density (5) which happens to be the analogue of the
Laplacian operator of differential geometry.

Similarly, we obtain the following bosonic transformations
\begin{eqnarray}
&& s_{\bar w} B_{\mu\nu} = - (\partial_\mu \bar {\cal B}_\nu - \partial_\nu \bar {\cal B}_\mu 
+ \frac{1}{2}
\varepsilon_{\mu\nu\eta\kappa} \partial^\eta \bar B^\kappa), 
\quad s_{\bar w} C_\mu = - \partial_\mu \lambda,
\nonumber\\
&& s_{\bar w} \bar C_\mu = - \partial_\mu \rho, \quad 
s_{\bar w} [\rho, \lambda, \phi_1, \phi_2, \beta, \bar\beta,
\bar B_\mu, \bar {\cal B}_\mu ] = 0, 
\end{eqnarray}
from the anticommutator $\{ s_{ab}, s_{ad} \}$. This bosonic transformation
is a {\it symmetry} transformation for the Lagrangian density (6) because
\begin{eqnarray}
s_{\bar w} {\cal L}_{(\bar B, \bar {\cal B})} &=& \partial_\mu \Bigl [\lambda (\partial^\mu \rho) - 
(\partial^\mu \lambda) \rho - \bar B^\kappa \partial^\mu \bar {\cal B}_\kappa \nonumber\\
 &+& \bar {\cal B}^\kappa \partial^\mu \bar B_\kappa - \bar {\cal B}^\mu (\partial \cdot \bar B) 
+ \bar B^\mu (\partial \cdot \bar {\cal B}) \Bigr ]. 
\end{eqnarray}
Thus, we have obtained a couple of bosonic symmetry transformations $s_w$ and $s_{\bar w}$
(cf. (13) and (15))
for the Lagrangian densities (5) and (6) which are derived from the {\it basic}
nilpotent symmetry transformations of the theory.

\section{Algebraic structures of the symmetry transformations and their relevance}

The appropriate Lagrangian densities of our present 4D Abelian 2-form gauge
theory are the ones given in (5) and (6). It has been demonstrated that the
Lagrangian density (5) is endowed with the BRST ($s_b$), co-BRST ($s_d$) and a
bosonic ($s_w = \{s_b, s_d \}$) symmetry transformations. The operator form
of these symmetry transformations are as follows
\begin{eqnarray}
&& s_b^2 = 0, \qquad s_d^2 \equiv \frac{1}{2} \{ s_{d}, s_{d} \}= 0, 
\qquad s_w = \{ s_b, s_d \}, \nonumber\\
&& [s_w, s_b ] = 0, \;\qquad [s_w, s_d ] = 0, \;\qquad s_w = (s_b + s_d)^2,
\end{eqnarray}
where it is understood that these algebraic relations act on any
arbitrary (i.e. the generic) field of the Lagrangian density (5). For instance,
the operator relation $s_w = \{ s_b, s_d \}$ implies that
$s_w \Omega_1 = \{ s_b, s_d \} \Omega_1$ where $\Omega_1 (= B_{\mu\nu},
B_\mu, {\cal B}_\mu,$ $ \phi_1, \phi_2, \beta, \bar \beta, C_\mu, \bar C_\mu,
\rho, \lambda)$  is the generic field of (5).

Exactly in a similar manner, we note that the anti-BRST ($s_{ab}$), anti-co-BRST
($s_{ad}$) and a bosonic ($s_{\bar w} = \{ s_{ab}, s_{ad} \}$) symmetry transformations
are respected by the Lagrangian density (6) of our present 2-form gauge theory. It
can be checked explicitly that these transformations, in the operator form, obey
the same kind of algebra as (17). In fact, this ensuing algebra is 
\begin{eqnarray}
&& s_{ab}^2 \equiv \frac{1}{2} \{ s_{ab}, s_{ab} \} = 0, 
\qquad s_{ad}^2 = 0, \quad s_{\bar w} = \{ s_{ab}, s_{ad} \}, \nonumber\\
&& [s_{\bar w}, s_{ab} ] = 0, \;\qquad [s_{\bar w}, s_{ad} ] = 0, \;\qquad 
s_{\bar w} = (s_{ab} + s_{ad})^2,
\end{eqnarray}
where it should be kept in mind that the above operator form of 
transformations act on the generic field $\Omega_2 (= B_{\mu\nu},
\bar B_\mu, \bar {\cal B}_\mu,$ $ \phi_1, \phi_2, \beta, \bar \beta, C_\mu, \bar C_\mu,
\rho, \lambda)$ of the Lagrangian density (6) of our present 2-form gauge theory.

The algebraic structures of (17) and (18) are exactly same as the algebra 
obeyed by the de Rham cohomological operators of differential geometry [11,12].
The following algebra of these celebrated operators 
\begin{eqnarray}
&&d^2 = 0, \;\qquad \;\delta^2 = 0, \;\qquad \;\Delta = \{ d, \delta \}, \nonumber\\
&& [\Delta, d ] = 0, \qquad [\Delta, \delta ] = 0, \qquad
\Delta = (d + \delta)^2,
\end{eqnarray}
captures the relationships amongst the exterior derivative $d = dx^\mu \partial_\mu$,
the co-exterior derivative $\delta = \pm * d *$ and the Laplacian operator $\Delta
= d \delta + \delta d$ which constitute the de Rham cohomological operators. Here
$*$ is the Hodge duality operation on a spacetime manifold without a boundary.

The anticommutativity of the following {\it basic} nilpotent transformations
\begin{eqnarray}
\{s_b, s_{ab} \} = 0, \quad \{s_b, s_{ad} \} = 0,  \quad
 \{s_d, s_{ab} \} = 0, \quad \{s_d, s_{ad} \} = 0,
\end{eqnarray}
is guaranteed only when the Curci-Ferrari type of restrictions (12) are exploited.
Furthermore, in addition to the transformations (7) and (9), the following 
off-shell nilpotent transformations on the auxiliary fields
\begin{eqnarray}
&& s_b \bar {\cal B}_\mu = 0, \;\quad \; s_d \bar B_\mu = 0, \;\quad \;s_{ad} B_\mu = 0, \;\quad\;
s_d \bar {\cal B}_\mu = - \partial_\mu \rho, \nonumber\\
&& s_b \bar B_\mu = - \partial_\mu \lambda, \quad s_{ab} B_\mu = \partial_\mu \rho, \quad
s_{ad} {\cal B}_\mu = \partial_\mu \lambda, \quad s_{ab} {\cal B}_\mu = 0,
\end{eqnarray}
are to be taken into consideration for the full proof of the anticommutativity. We re-emphasize that
the restrictions (12) are (anti-)BRST as well as (anti-)co-BRST
invariant quantities as can be checked by using (7), (9) and (21).

 \section{Conclusions}
 
The appropriate set of coupled and equivalent Lagrangian densities (cf. (4)) for the 4D
free Abelian 2-form gauge theory were proposed in [10] which were endowed with
the off-shell nilpotent and anticommuting (anti-)BRST symmetry transformations.
The central theme of our present investigation was to generalize the Lagrangian
densities of [10] so as to obtain off-shell nilpotent (anti-)co-BRST 
transformations {\it together} with the (anti-)BRST symmetry transformations
of [10]. We have achieved this goal in the form of the Lagrangian densities
(5) and (6) of our present investigation.

To accomplish the above objective, we have introduced a pair of Lorentz vector
auxiliary fields (i.e. ${\cal B}_\mu, \bar {\cal B}_\mu$)  and a massless
($ \Box \phi_2 = 0$) scalar field $\phi_2$ to generalize the Lagrangian
densities (cf. equation (4)) of our earlier work [10]. It turns out that
a pair of Curci-Ferrari type of restrictions (cf. (12)) are required to
obtain the absolute anticommutativity of the (anti-)BRST as well as (anti-)co-BRST symmetry 
transformations {\it together} for the Abelian 2-form gauge theory. These restrictions
are found to be invariant under the (anti-)BRST as well as (anti-)co-BRST
symmetry transformations as can be checked by exploiting the
transformations (7), (9) and (21).

The above four basic nilpotent transformations are the {\it symmetry} transformations
for the Lagrangian densities (5) and (6) of our present 2-form theory and they
correspond to the exterior and co-exterior derivatives of the differential
geometry. The anticommutator of the BRST and co-BRST transformations
produces a bosonic symmetry transformation which is the analogue
of the Laplacian operator for the Lagrangian density (5). In a similar fashion, the 
anticommutator of the anti-BRST and anti-co-BRST transformations leads to
the derivation of a bosonic symmetry transformation that corresponds to the Laplacian
operator for the Lagrangian density (6).

One of the Curci-Ferrari type of restriction in (12) has been shown, in our work [7],
to have a deep connection with the concept of gerbes which have become very active area
of research in theoretical high energy physics (see, e.g. [13,14]). It is a challenging
problem for us to find out the meaning of the other restriction (i.e. ${\cal B}_\mu -
\bar {\cal B}_\mu + \partial_\mu \phi_2 = 0$) in the language of gerbes.

We have discussed, in our present endeavour, only the continuous symmetry transformations
that imbibe the algebraic structure of the de Rham cohomological operators. This exercise 
is our preparation to accomplish our main goal of proving the present Abelian 2-form 
gauge theory to be a field theoretic model for the Hodge theory. We have achieved this goal
in our recent paper [15]. Right now, the Hamiltonian analysis of our present
model is being investigated. Our new results would be reported elsewhere [16].\\
 
\noindent
{\bf Acknowledgement:}
Financial support from the
Department of Science and Technology (DST), Government of India,  
under the SERC project sanction grant No: - SR/S2/HEP-23/2006, is gratefully acknowledged.

\end{document}